\documentclass[aps,pre,reprint,10pt,superscriptaddress,showpacs]{revtex4-1}
\usepackage{amsfonts}
\usepackage{amsmath}
\usepackage{amssymb}
\usepackage{graphicx}%
\usepackage{color}
\begin{document}

\title{ Electrostatic solitary waves in dusty pair-ion plasmas }
\author{A. P. Misra}
\email{apmisra@visva-bharati.ac.in; apmisra@gmail.com}
\affiliation{Department of Mathematics, Siksha Bhavana, Visva-Bharati University, Santiniketan-731 235, West Bengal, India}
\author{N. C. Adhikary}
\email{nirab$_$iasst@yahoo.co.in}
\affiliation{Physical Sciences Division, Institute of Advanced Study in Science and Technology, Vigyan Path, Paschim Boragaon, Garchuk, Guwahati-781035, Assam, India}
\begin{abstract}
The  propagation of  electrostatic  waves in an unmagnetized collisionless pair-ion plasma with immobile  positively charged dusts is studied for both large- and small-amplitude perturbations.  Using a two-fluid model for pair-ions, it is shown that  there appear two linear ion  modes, namely the ``fast" and ``slow" waves in dusty pair-ion plasmas. The properties of these wave modes are studied  with different mass $(m)$ and temperature $(T)$ ratios of negative to positive   ions, as well as the effects of immobile charged dusts $(\delta)$. For large-amplitude waves, the pseudopotential approach is performed, whereas the     standard reductive perturbation technique (RPT) is used to study the small-amplitude  Korteweg-de Vries (KdV) solitons.    The    profiles of the pseudopotential, the large amplitude solitons as well as the dynamical evolution of KdV solitons are numerically studied   with the system parameters as above.   It is found that the pair-ion plasmas with positively charged  dusts support  the  propagation of solitary waves (SWs) with only the negative potential.  The results may be useful for the excitation of  SWs in laboratory dusty pair-ion  plasmas, electron-free industrial plasmas as well as for observation in space plasmas where electron density is negligibly small compared to that of negative ions.
\end{abstract}
\maketitle
\section{Introduction} 
A  typical  plasma consisting of electrons and positive ions essentially causes temporal as well as spatial variations of collective phenomena due to large-mass difference between these particles. This asymmetric diversity of collective  plasma phenomena can, however, be nullified in pair-ion plasmas consisting of positive and negative ions with equal mass. In the latter, the space-time parity can be maintained because of the same mobility of the particles under electromagnetic forces. Such pair-ion plasmas have been generated in the laboratory, and three kinds of electrostatic modes have been experimentally observed to propagate along magnetic-field lines  in paired fullerene-ion   plasmas \cite{pair-ion-experiment}. Furthermore, in many industries such as integrated-circuit fabrication, there requires a plasma source having no energetic electrons in the plasma, since the deposited film is strongly damaged by a high-energy electron.  For this purpose,  a radio-frequency plasma source has also been developed \cite{rf-plasma}. 

On the other hand, \textit{in situ} measurements of charged particles in the polar mesosphere under nighttime conditions revealed that there exist positively charged nanoparticles \cite{in-situ-Rapp}.  Such particles have been observed in a region   dominated by both positive and negative ions, and very few percentage of electrons. It was also clarified that the positive charge of these dust particles    is due to the dominant charging effects of lighter positive ions compared to the heavier negative ions, and the presence of a very small number of electrons. Furthermore, in an experiment,   it has been   investigated that dust particles injected  in pair-ion plasmas can become positively charged  when the number density of negative ions greatly exceeds ($\gtrsim500$) that of the electrons \cite{Kim-Merlino}. In space environments, the possible role of negative ions has been discussed, and it has been found that  dusts can be positively charged if there is a sufficient number density of heavy negative ions (with mass $\gtrsim300$ amu) \cite{in-situ-Rapp}. Thus, in laboratory and space environments, pair-ion plasmas with positively charged dusts and no high-energy electrons may not be unubiquitous. So,   collective plasma oscillations and nonlinear properties of electrostatic as well as electromagnetic waves in these pair-ion plasmas are worth investigating. 

To mention few, the nonlinear propagation of solitary waves (SWs) and shocks     in dusty plasmas have been widely studied   for understanding the electrostatic disturbances in space plasma environments  as well as in laboratory plasma devices \cite{space-1,space-2,lab-3,lab-4,Nakamura-Sarma}.  It has been   shown that  charged dust grains can drastically modify the existing response of electrostatic wave spectra in plasmas depending upon whether the charged dusts are considered to be static or mobile \cite{lab-3,lab-4,5,6,7,8,9,10,11,12}. Recently, there has been a growing interest in investigating the properties of electrostatic waves in pair-ion plasmas (see, e.g., Refs. \cite{pair-ion1,pair-ion2}). However, to our knowledge, no detailed theory has been made  to study the electrostatic small- as well as large-amplitude waves in dusty pair-ion plasmas with positively charged dusts.
 
So,  our purpose   is to investigate the propagation characteristics of electrostatic large- as well as small-amplitude waves in unmagnetized collisionless dusty pair-ion plasma with positively charged dusts. We show that there exist two modes: ``fast" and ``slow" waves, the properties of which are studied.   We use the pseudopotential approach to study the properties of large-amplitude waves, whereas the reductive perturbation technique is used to investigate  Korteweg-de Vries (KdV) solitons.  The present work thus generalizes and extends the previous works, e.g. Ref. \cite{pair-ion3} to include the effects of positively charged dusts, and different mass    as well as thermal pressures of ions. It is shown that in   dusty pair-ion plasmas, SWs exist with  only  the negative potential.  
\section{Basic equations}
We consider the one-dimensional propagation of electrostatic waves in an unmagnetized collisionless dusty plasma  consisting of singly charged  adiabatic positive and negative ions, and immobile positively charged dusts.  We do not consider the dynamics of charged dusts as they are too heavy   to move on the time scale of the ion-acoustic waves. However, dusts can affect the  wave dispersion and nonlinearity. Furthermore, we assume that  the negative ions are heavier than the positive ions, and negative ion number density is much larger than that of electrons so that  dusts become positively charged \cite{Kim-Merlino}.   Thus, the dominant higher mobility species in the plasma are  the positive ions.       The immobile dust particles carry some charges so as to maintain the overall charge neutrality condition given by
\begin{equation}
n_{n0}=n_{p0}+z_dn_{d0}, \label{charge-neutrality-1}
\end{equation} 
where $n_{j0}$ is the unperturbed number density of charged species $j$ ($j=p$, $n$, $d$, respectively, stand for  positive ions, negative ions and static positively charged dusts), $z_d$ $(>0)$ is the unperturbed dust charge state.  The condition  \eqref{charge-neutrality-1}, in dimensionless form, becomes
\begin{equation}
\mu+\delta=1, \label{charge-neutrality-2}
\end{equation}
where $\mu=n_{p0}/n_{n0}$ and  $\delta=z_dn_{d0}/n_{n0}$ are the density ratios. 
 The basic equations to describe the dynamics of positive and negative ions  in one space dimension are
\begin{equation}
\frac{\partial n_j}{\partial t}+\frac{\partial}{\partial x}(n_j v_j)=0, \label{cont-eq}
\end{equation}
\begin{equation}
\left(\frac{\partial}{\partial t}+v_j\frac{\partial}{\partial x}\right)v_j=-\frac{q_j}{m_j}\frac{\partial \phi}{\partial x}-\frac{3k_BT_j}{2m_jn^2_{j0}}\frac{\partial n^2_j}{\partial x}, \label{moment-eq}
\end{equation}
\begin{equation}
\frac{\partial^2 \phi}{\partial x^2}=4\pi e\left(n_n-n_p-z_dn_{d0}\right),\label{poisson-eq}
\end{equation}
where  $n_j$, $v_j$, and $m_j$, respectively, denote the number density, velocity, and mass of $j$-species particles. Furthermore, $q_{p,n}=\pm e$,  where $e$ is the elementary charge. Also, $\phi$ is the electrostatic potential, $k_B$ is the Boltzmann constant, and $T_j$ is the particle's thermodynamic temperature. In Eq. \eqref{moment-eq},    we have used the adiabatic pressure law: $P_j/P_{j0}=(n_j/n_{j0})^3$ with $P_{j0}=n_{j0}k_B T_j$ for each  ion-species $(j=p,n)$.  The adiabatic index $\gamma=3$ $[=(2+D)/D$, $D$ being the number of degrees of freedom] is used for one-dimensional geometry $(D=1)$ of the system. We will later see that in the long-wavelength limit, the phase velocity of the fast wave is much greater than the positive-ion thermal velocity. This certainly justifies our assumption  of adiabaticity that thermal conduction cannot keep up with the moving wave
front.
  
Next, we normalize the physical quantities according to
$\phi\rightarrow e\phi/k_BT_p$, $n_j\rightarrow n_j/n_{j0}$, $v_j\rightarrow v_j/c_s$, where $c_s=\sqrt{k_BT_p/m_n}=\omega_{pn}\lambda_D$ is the ion-acoustic speed with $\omega_{pn}=\sqrt{4\pi n_{n0}e^2}/m_n$ and $\lambda_D=\sqrt{k_BT_p/4\pi n_{n0}e^2}$ denoting, respectively, the negative-ion plasma frequency and the Debye length. The space and time variables are normalized by  $\lambda_D$  and  $\omega^{-1}_{pn}$  respectively. Thus, from Eqs. \eqref{cont-eq}-\eqref{poisson-eq}, the normalized set of equations is obtained as: 
\begin{equation}
\frac{\partial n_j}{\partial t}+\frac{\partial}{\partial x}(n_j v_j)=0, \label{Ncont-eq}
\end{equation}
\begin{equation}
\left(\frac{\partial}{\partial t}+v_j\frac{\partial}{\partial x}\right)v_j=-\frac{m_n}{m_j}\left(\pm\frac{\partial \phi}{\partial x}+\frac{3}{2}\frac{T_j}{T_p}\frac{\partial n^2_j}{\partial x}\right), \label{Nmoment-eq}
\end{equation}
\begin{equation}
\frac{\partial^2 \phi}{\partial x^2}=n_n-\mu n_p-\delta,\label{Npoisson-eq}
\end{equation}
where  the upper (lower) sign of $\pm$ on the right-hand side of Eq. \eqref{Nmoment-eq}  corresponds to positive (negative) ions. For convenience and to be used later, we denote $T=T_n/T_p$  and $m=m_n/m_p$. 
\section{Dispersion relation: ion-wave modes}
In order to figure out the number of linear wave modes and to study their properties,    we  assume the perturbed  quantities to vary as $\sim\exp(ikx-i\omega t)$, i.e., in the form of oscillations with wave frequency  $\omega$ and wave number $k$. Thus, Fourier analyzing the  Eqs. \eqref{Ncont-eq}-\eqref{Npoisson-eq}, we obtain the following  {dispersion relation:
\begin{equation}
\frac{m\mu}{\omega^2-3mk^2}+\frac{1}{\omega^2-3Tk^2}=1,\label{Dispersion-R}
\end{equation}
in which $\omega$ and $k$ are normalized by $\omega_{pn}$ and $\lambda^{-1}_D$ respectively. The first and second terms on the left-hand side of Eq. \eqref{Dispersion-R} are,  respectively, due to the presence of positive- and negative-ion species, whereas the  constant term $1$ on the right-hand side appears due to the effect of charge separation of the species (deviation from quasineutrality). 
The dispersion  Eq. \eqref{Dispersion-R}  admits two  solutions of $\omega^2$ given by
\begin{equation}
 \omega^2=\frac{1}{2}\left(\omega_1^2\pm\sqrt{\omega_1^4-\omega_2^4}\right),\label{Disp1}
\end{equation}
where $\omega_1^2=1+m\mu+3(T+m)k^2$ and $\omega_2^2=\sqrt{12m\left[1+T\left(\mu+3k^2\right)\right]k^2}$. In order that these    solutions  are real, the discriminant
 \begin{eqnarray}
 &&\omega_1^4-\omega_2^4=9k^4(m-T)^2+6k^2(m-T)(m\mu-1)\notag\\
 &&\hskip50pt+(1+m\mu)^2 \label{dis-criminant}
\end{eqnarray}     
must be nonnegative. We find that  the discriminant is positive for all $k$, when $m>T$, $m>1$, and $\mu<1$ are satisfied.   
 In Eq. \eqref{Disp1}, the positive and negative signs, respectively, correspond to ``fast" and ``slow" wave modes in pair-ion plasmas. Recalling Eq. \eqref{Dispersion-R} in dimensional form, we find that the effect of dispersion due to charge separation can not be neglected, because otherwise,  for the ``fast" waves of which the phase speed is  much larger than the ion-thermal speed,  the sum of the  two   positive terms becomes zero, which is inadmissible. Furthermore, in order to avoid the wave damping due to the resonance with  either the positive or negative ions, the phase speed is to be much larger than the negative ion thermal speed and much lower than the positive ion thermal speed.    In particular, in absence of the charged dusts, and if ions, having the same mass, are in  isothermal equilibrium,   one can easily recover the dispersion equation (6) in Ref. \cite{pair-ion3}. Furthermore, in particular, for $m=T=\mu=1$, i.e., in dust-free plasmas in which ions   have the same mass and  temperature, the two ion-modes, corresponding to positive and negative signs in Eq. \eqref{Disp1}, reduce, in the long-wavelength limit $(k\rightarrow0)$, to $\omega^2=2+3k^2$ and $\omega^2=3k^2$, or, in dimensional form, $\omega^2=2\omega_{pn}^2+3c_s^2k^2$ and $\omega^2=3c_s^2k^2$. The former is the usual ion-plasma wave, which propagates with a frequency  greater than   $\omega_{pn}$ (even in absence of the thermal pressure), whereas, the latter propagates forward in the manner of a sound wave in which the electrostatic oscillation disappears. It has been observed in experiments that slow-wave modes do not favor the formation of solitons, whereas   the fast  modes  may propagate as   solitary waves  due to nice balance of the dispersive and nonlinear effects \cite{experiment-negative-ion,experiment-Wong}. In the next sections, our attempt will be to study whether this ion-plasma wave (fast mode) propagates as solitary waves. However, before going into the detail investigation for the nonlinear propagation, we first analyze numerically the properties of the ``fast" and ``slow" wave modes as follows:

\begin{figure*}
\includegraphics[height=4in,width=6in]{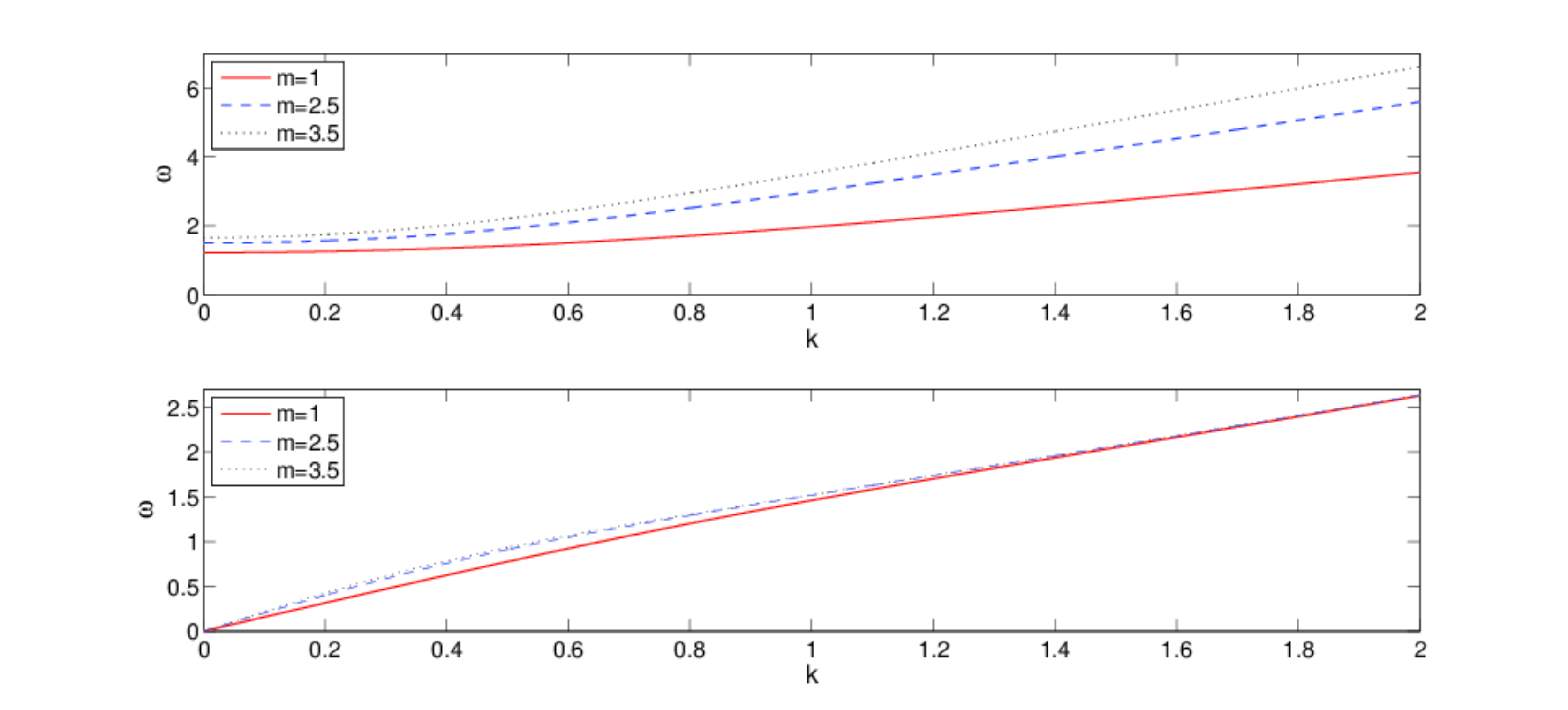}  
\caption{(Color online) The plots of  the wave frequency $\omega$ versus the wave number $k$ obtained as a  solution of the dispersion relation [Eq. \eqref{Dispersion-R}] for different values of $m$ as indicated in the figure. The upper (lower) panel is for the fast (slow) wave modes. The solid, dashed and dotted lines correspond to $m=1$, $2.5$ and $3.5$ respectively. The other parameter values are  $T=\delta=0.5$.} 
\end{figure*}
\begin{figure*}
\includegraphics[height=4in,width=6in]{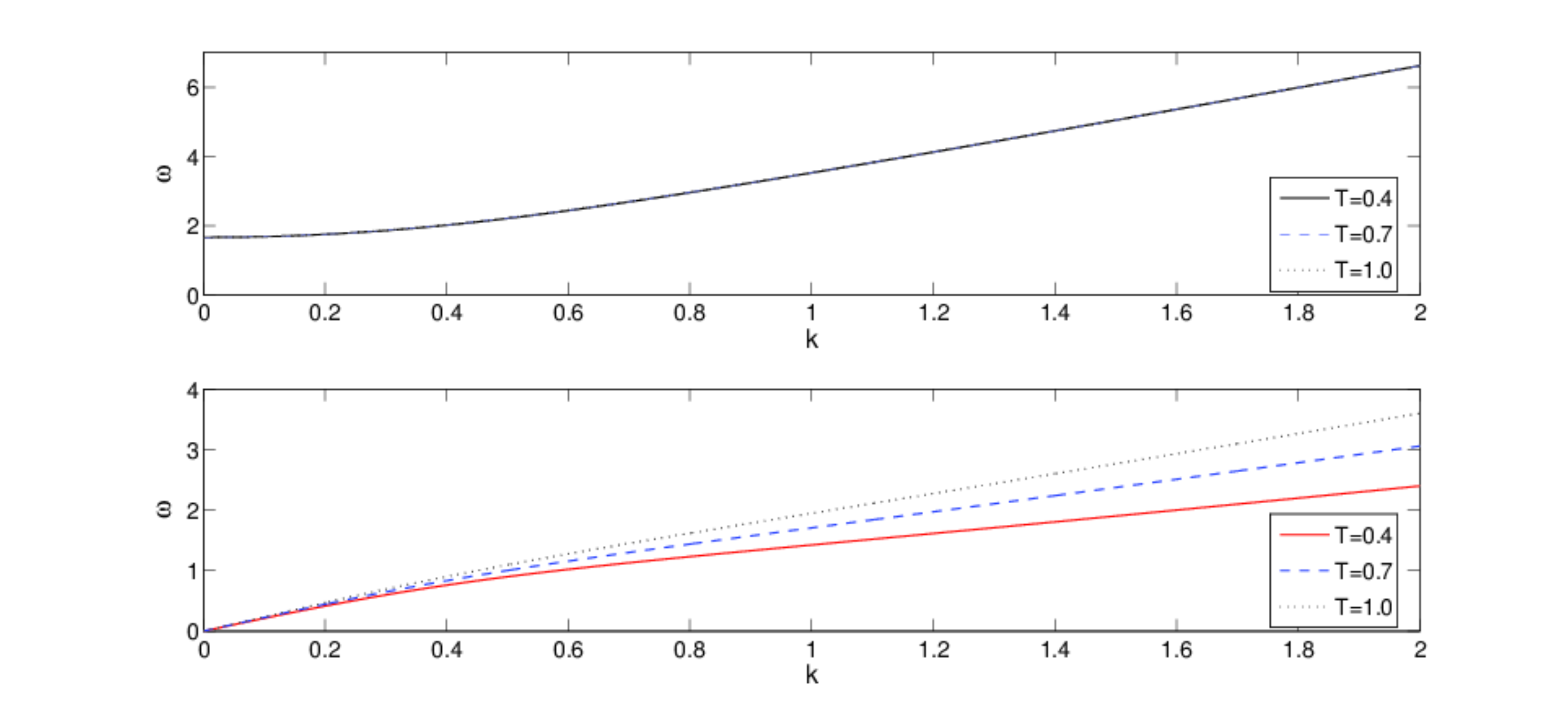}  
\caption{(Color online) The plots of  the wave frequency $\omega$ versus the wave number $k$ obtained as a  solution of the dispersion relation [Eq. \eqref{Dispersion-R}] for different values of $T$ as indicated in the figure. The upper (lower) panel is for the fast (slow) wave modes. The solid, dashed and dotted lines correspond to $T=0.4$, $0.7$ and $1.0$ respectively. The other parameter values are  $m=3.5$ and $\delta=0.5$.} 
\end{figure*}
\begin{figure*}
\includegraphics[height=4in,width=6in]{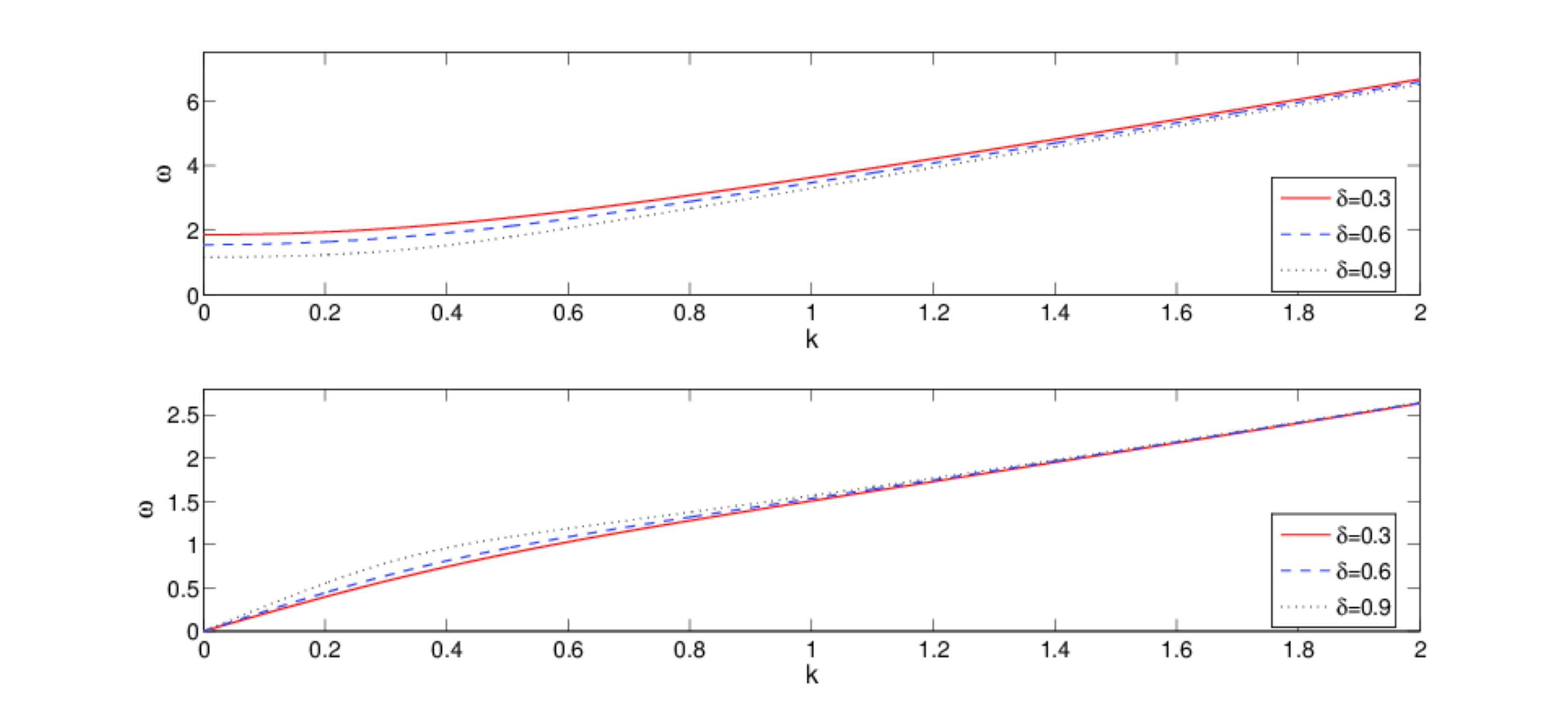}  
\caption{(Color online) The plots of  the wave frequency $\omega$ versus the wave number $k$ obtained as a  solution of the dispersion relation [Eq. \eqref{Dispersion-R}] for different values of $\delta$ as indicated in the figure. The upper (lower) panel is for the fast (slow) wave modes. The solid, dashed and dotted lines correspond to $\delta=0.3$, $0.6$ and $0.9$ respectively. The other parameter values are  $T=0.5$ and $m=3.5$.} 
\end{figure*}
Figures 1 to 3 show the plots of the frequencies of the fast- (upper panel) and slow-wave (lower panel) modes    with the variations of the wave number $k$ for different sets of parameter values of $m$, $T$ and $\delta$ as indicated in the figures. From Fig. 1, it is evident  that both the fast- and slow-wave frequencies increase with $k$ as well as with higher values of the mass ratio $m$. In contrast to the fast mode, the slow-wave frequency remains almost unaltered for values of $m\gtrsim2.5$,  as well as in the limits of the long-wavelength, $k\rightarrow0$ and short-wavelength, $k\gtrsim1.4$. In the former limit,  though the frequency of the fast mode   bears almost a constant value,  the slow-wave frequency approaches to zero.  In all the figures 1 to 3, the wave frequency is found to increase with the wave number $k$ implying that the wave phase speed exceeds that of the ion-sound wave. Figure 2 shows the similar profiles as in Fig. 1, but for different values  of the temperature ratio $T$. It is found that $T$, however, does not affect the frequency of the fast-wave mode (see the upper panel of Fig. 2), whereas the slow-wave frequency increases with increasing values of $T$. In the latter, $\omega$ becomes larger with higher values of $k\gtrsim1$. Figure 3 shows that in contrast to the slow-wave modes, the effect of positively charged dusts is to decrease the wave frequency of the fast-wave modes. It is also seen that  for $k>1$, charged dusts  have very small effect on the wave modes.

\section{Large amplitude solutions: Pseudopotential approach}
Here, we assume  the perturbations to vary in a moving frame of reference $\zeta=x-Mt$, where $M$ is the nonlinear wave speed normalized by  $c_s$. {We also use the boundary conditions, namely, $n_j\rightarrow1$, $v_j\rightarrow0$, $\phi\rightarrow0$ as $\zeta\rightarrow\pm\infty$.} Then from the continuity equation \eqref{Ncont-eq} {we obtain the following expressions for the positive and negative ion fluid velocities $(j=p,n)$:}
\begin{equation}
v_j=M\left(1-\frac{1}{n_j}\right) \label{cont1}
\end{equation}
Next, from Eq. \eqref{Nmoment-eq} and Eq. \eqref{cont1}, after eliminating $v_j$, {we obtain bi-quadratic equations for the densities $n_j$ which have the following solutions: 
\begin{equation}
n_{p\pm}=\frac{1}{\sqrt{3m}}\left[f_{p}(\phi)\pm\sqrt{f_{p}^2(\phi)-\alpha_p^2} \right]^{1/2}, \label{density-p}
\end{equation}
\begin{equation}
n_{n\pm}=\frac{1}{\sqrt{3T}}\left[f_{n}(\phi)\pm\sqrt{f_{n}^2(\phi)-\alpha_n^2} \right]^{1/2}, \label{density-n}
\end{equation}
where $n_{j\pm}$ are the values of the number densities $n_j$ corresponding to the   $\pm$ signs. Also,  $f_p(\phi)=3m/2+M^2/2-m\phi$,  $f_n(\phi)=3T/2+M^2/2+\phi$,$\alpha_p=\sqrt{3m}M$, and $\alpha_n=\sqrt{3T}M$}. Equations \eqref{density-p} and \eqref{density-n} give four possible combinations for the number densities, namely, $n_{p+}$, $n_{n+}$; $n_{p-}$, $n_{n-}$, $n_{p+}$, $n_{n-}$ and $n_{p-}$, $n_{n+}$. We will later see that only one of them will favor the propagation of large-amplitude solitary waves. From Eqs. \eqref{density-p} and \eqref{density-n} it is clear that for   real values of the densities,  the conditions  {$f_{j}(\phi)>0$ and $f_j(\phi)\geq \alpha_j$ must be satisfied. These lead to the following bounds for the electrostatic potential: 
\begin{equation}
\phi_l<\phi<\phi_u, \label{phi-range}
\end{equation}
where  $\phi_l=-\left(M-\sqrt{3T}\right)^2/2<0$ and $\phi_u=\left(M-\sqrt{3m}\right)^2/2m>0$, implying that $\phi$ has the range from negative to positive values.} 
Next, from the Poisson equation \eqref{Npoisson-eq}, after integrating with respect to $\zeta$ and using the above boundary conditions, we obtain the following energy-like equation for an oscillating particle of unit mass at the pseudoposition $\phi$ and pseudotime $\zeta$:
\begin{equation}
\frac{1}{2}\left(\frac{d\phi}{d\zeta}\right)^2+V\left(m,\mu,M,T,\phi\right)=0,\label{energy-equation}
\end{equation}
where the pseudopotential $V$ is given by
{
\begin{eqnarray}
V=(1-\mu)\phi-\frac{2\mu}{(3m)^{3/2}}\left[g_{p\pm}(\phi)-g_{p\pm}(0)\right] \notag\\
-\frac{2}{3\sqrt{3T}}\left[g_{n\pm}(\phi)-g_{n\pm}(0)\right],\label{V-pseudo}
\end{eqnarray}
with 
\begin{equation}
g_{j\pm}(\phi)=h_{j\pm}(\phi)\left(2f_{j}(\phi)\mp\sqrt{f_{j}^2(\phi)-\alpha_j^2}\right), \label{g-function}
\end{equation} and $h_{j\pm}(\phi)=\left(f_{j}(\phi)\pm\sqrt{f_{j}^2(\phi)-\alpha_j^2}\right)^{1/2}$. Here, $g_{j\pm}(0)$ is the value of $g_{j\pm}(\phi)$ at $\phi=0$. The pseudopotential $V$ and hence the relevant expressions, to be shown shortly, can have four different forms depending on the consideration of the densities $n_{j\pm}$  given by Eqs. \eqref{density-p} and \eqref{density-n}.  However, we will show that only one of them   corresponds to the existence of solitary waves or double layers.  The energy-like equation \eqref{energy-equation}, which describes the evolution of arbitrary amplitude electrostatic perturbations, can also be obtained following Refs. \cite{sagdeev1,sagdeev2}}. The conditions for which the perturbations may propagate as solitary waves or double layers can be discussed   as follows:\\\\
\textbf{Condition 1:} $V(\phi)=0$ at $\phi=0$. This can be easily verified from Eq. \eqref{V-pseudo} by substituting $\phi=0$. Also, $V(\phi)<0$ for $\phi_l<\phi<\phi_u$. This condition will be examined numerically later.\\\\
\textbf{Condition 2:} $\frac{dV}{d\phi}=0$ at $\phi=0$. From Eq. \eqref{V-pseudo}  we have  {
\begin{equation}
\left(\frac{dV}{d\phi}\right)_{\phi=0}=1-\mu+\frac{\mu h_{p\pm}(0)}{\sqrt{3m}}-\frac{h_{n\pm}(0)}{\sqrt{3T}}, \label{dv-dphi}
\end{equation} 
where \begin{equation}
h_{p\pm}(0)=\left\lbrace \begin{array}{lll}
 \left(\sqrt{3m},M\right)  & \text{if}& M^2\leq3m, \\
 \left(M,\sqrt{3m}\right)  & \text{if}& M^2>3m,
 \end{array}      \right. \label{dv-dphi-p}
  \end{equation}
and
\begin{equation}
 h_{n\pm}(0)=\left\lbrace \begin{array}{lll}
 \left(\sqrt{3T},M\right)  & \text{if}& M^2\leq3T, \\
 \left(M,\sqrt{3T}\right)  & \text{if}& M^2>3T,
 \end{array}      \right. \label{dv-dphi-n}
  \end{equation}}
in which the first and second terms in the parentheses are corresponding to the positive and negative signs of $\pm$. Thus, from Eqs. \eqref{dv-dphi-p} and \eqref{dv-dphi-n} we find that {
\begin{itemize}
\item{If $M\leq M_{\text{min}}\equiv\text{min}\lbrace\sqrt{3m},\sqrt{3T}\rbrace$, then the condition $\frac{dV}{d\phi}=0$ at $\phi=0$ is satisfied for the expression [Eq. \eqref{dv-dphi}] corresponding to $h_{j+}(0)$ (i.e., corresponding to the number densities $n_{j+}$).  However, for the expression corresponding to $h_{j-}(0)$ (i.e., corresponding to the number densities $n_{j-}$), the same condition is satisfied when $M=M_0\equiv(1-\mu)\sqrt{3mT}/(\sqrt{m}-\mu\sqrt{T})$. Typically, for laboratory and space plasmas\cite{APM1}, $m\geq1$ and $T\leq1$.  So,  $M_{\text{min}}=\sqrt{3T}$, and $M=M_0\leq M_{\text{min}}=\sqrt{3T}$ holds for $m\geq T$ together with $\mu<1$ for positively charged dusts.}
\item{If  $M>M_{\text{max}}\equiv\text{max}\lbrace\sqrt{3m},\sqrt{3T}\rbrace$, i.e., if $M>\sqrt{3m}$ for $m\geq T$, then $\frac{dV}{d\phi}=0$ at $\phi=0$ is satisfied  for the expression corresponding to $h_{j-}(0)$, i.e., corresponding to the densities $n_{j-}$. However,  for $h_{j+}(0)$ or for the densities $n_{j+}$, the same is satisfied for $M=M_0$. But, in this case, $M_0\not>\sqrt{3m}$, otherwise, we would have $T>m$, which contradicts the above consideration  $m\geq T$.}
\item{When $\sqrt{3T}<M<\sqrt{3m}$ holds for $m\geq T$ together with $\mu<1$, $m\geq1$ and $T\leq1$, the condition $\frac{dV}{d\phi}=0$ at $\phi=0$ is satisfied corresponding to $h_{p+}(0)$ and $h_{n-}(0)$, i.e., for the number densities $n_{p+}$   and  $n_{n-}$.}   
\end{itemize}
The other cases may not be of interest for the parameter regimes mentioned above.
   Thus, depending on the consideration of the solutions of the number densities $n_{j\pm}$, and hence the expressions for $V(\phi)$ and its derivatives,  the condition $\frac{dV}{d\phi}=0$ at $\phi=0$ may be satisfied for a certain range of values of  $M$, namely, $M<\sqrt{3T}$, $M>\sqrt{3m}$ or $\sqrt{3T}<M<\sqrt{3m}$.} \\\\
\textbf{Condition 3:} $\frac{d^2V}{d\phi^2}<0$ at $\phi=0$. From Eq. \eqref{V-pseudo}, we obtain  
\begin{equation}
\left(\frac{d^2V}{d\phi^2}\right)_{\phi=0}=\frac{\mp m\mu}{2\sqrt{3m}}\frac{ h_{p\pm}(0)}{\sqrt{f_p^2(0)-\alpha_p^2}}+\frac{\mp 1}{2\sqrt{3T}}\frac{ h_{n\pm}(0)}{\sqrt{f_n^2(0)-\alpha_n^2}},\label{d2vdphi2}
\end{equation}
where $\mp$ signs appear due to the expressions corresponding to $h_{j\pm}(0)$. From Eq. \eqref{d2vdphi2}, we find that (i) for $M<\sqrt{3T}$, $d^2V/d\phi^2$ at $\phi=0$ is always negative  when $d^2V/d\phi^2$ takes the form corresponding to $h_{j+}(0)$; (ii) for $M>\sqrt{3m}$,  $d^2V/d\phi^2$ at $\phi=0$ is always positive [when $d^2V/d\phi^2$ has the expression corresponding to $h_{j-}(0)$]; (iii) in the range $\sqrt{3T}<M<\sqrt{3m}$, $d^2V/d\phi^2$ at $\phi=0$ [corresponding to $h_{p+}(0)$ and $h_{p-}(0)$, i.e., for the number densities $n_{p+}$ and $n_{n-}$] is  negative for $M>\tilde{M}\equiv\sqrt{3m(1+\mu T)/(1+m\mu)}$. The latter is admissible for $m>T$. In this case, the range of $M$ is precisely $\text{max}\left\lbrace  \sqrt{3T},~\tilde{M}\right\rbrace<M<\sqrt{3m}$.  }\\\\ 
\textbf{Condition 4:} For a nonzero $\phi_m$, the relations $V(\phi_m)=0$ and $dV(\phi_m)/d\phi\gtrless 0$ are to be satisfied according to whether the solitary waves are compressive (with   $\phi>0$) or rarefactive (with   $\phi<0$). Here, $\phi_m$ $(\neq0)$ represents the amplitude of the solitary waves. However,   in addition, to the above conditions, if $dV(\phi_m)/d\phi$ vanishes instead of $dV(\phi_m)/d\phi\gtrless 0$, the perturbations may develop into  double layers. These conditions will also be examined numerically.  
\par  {From the above discussions we find that there may be three possible regions, for which all the above conditions for the existence of large amplitude solitary waves or double layers are satisfied, these are \\\\
(i)  \textbf{Region 1:} $0<M<\sqrt{3T}$ together with $m\gtrsim1$, $T\lesssim1$, $\mu<1$.  The expression for  $V$ and relevant others are corresponding to the number densities $n_{j+}$;\\\\
(ii) \textbf{Region 2:} $M>\sqrt{3m}$  together with $m\gtrsim1$, $T\lesssim1$, $\mu<1$.  Here, the expression for  $V$ and relevant others are corresponding to $n_{j-}$;\\\\ 
(iii)  \textbf{Region 3:} $\text{max}\left\lbrace  \sqrt{3T},~\tilde{M}\right\rbrace<M<\sqrt{3m}$ together with $m\gtrsim1$, $T\lesssim1$, $\mu<1$. In this case, the expression for  $V$ and relevant others are corresponding to $n_{p+}$ and $n_{n-}$.}\par  { In the following numerical investigation, we will
see that the existence of large-amplitude solitary waves may be possible only for the parameters in `Region 3'.}  It will also be shown that $dV(\phi_m)/d\phi\neq0$ except at $\phi=0$, i.e., the  double layers may not exist in the plasmas. 
\begin{figure*}
\includegraphics[height=4in,width=6in]{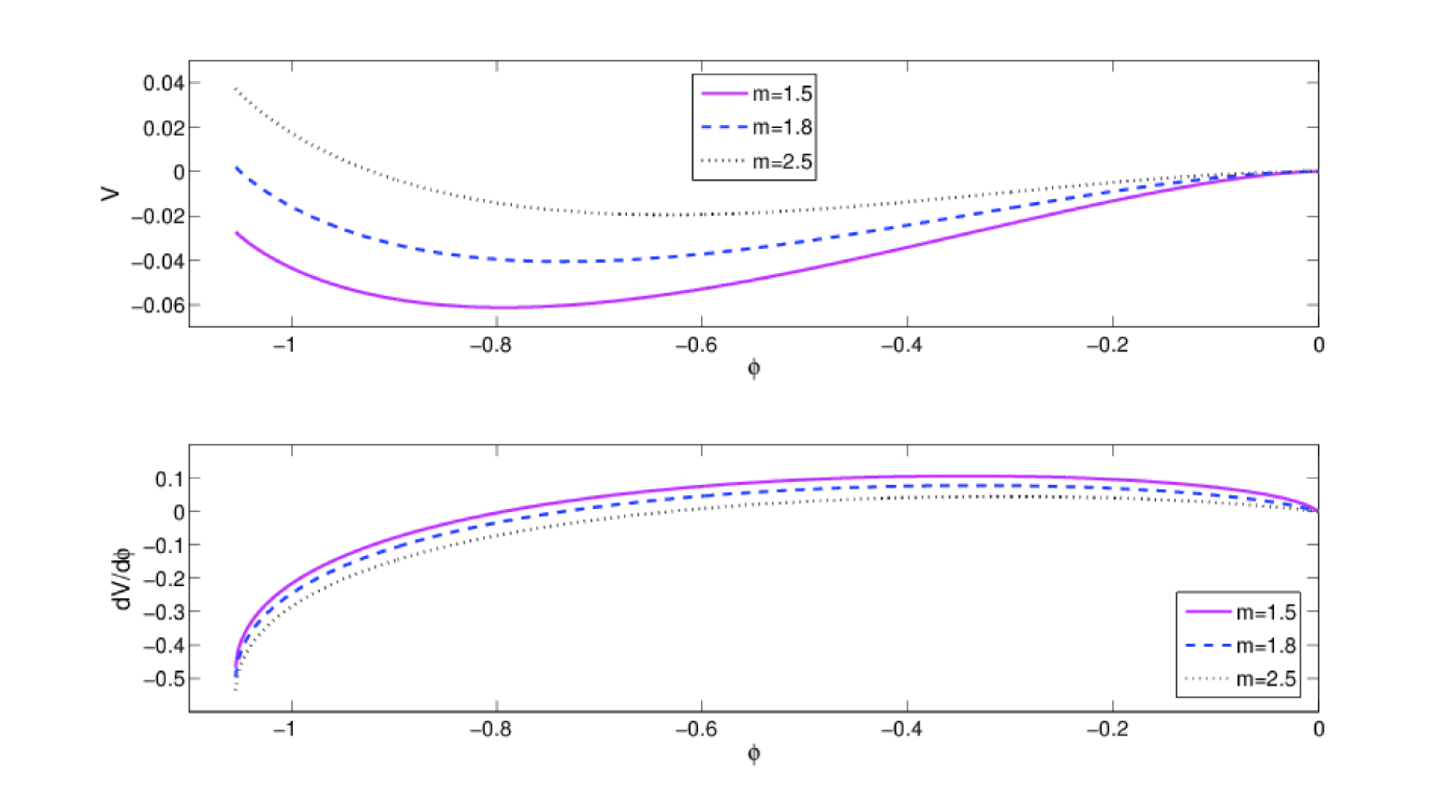}  
\caption{(Color online) The upper and lower panels, respectively, show the plots of the pseudopotential $V$ [Eq.\eqref{V-pseudo}] and its derivative $dV/d\phi$ [Eq. \eqref{dv-dphi}]   against the electrostatic potential $\phi$ for different values of the mass ratio $m$. The solid, dashed, dotted and dash-dotted lines correspond to  { $m=1.5$, $1.8$, and $2.5$  respectively. The other parameter (fixed) values are $M=2$, $T=0.1$ and $\delta=0.1$. We see that $V=0=dV/d\phi$ and $d^2V/d\phi^2<0$ (not shown) at $\phi=0$. Also, $V=0$ and $dV/d\phi<0$ at $\phi=-1.05$ (dashed line) and $-0.92$  (dotted line). The Conditions 1 to 4 are satisfied for $m\gtrsim1.8$ and for fixed parameters as above.  Furthermore, $dV(\phi_m)/d\phi\neq0$ except at $\phi=0$, implying the nonexistence of   double-layer solutions.}  } 
\end{figure*}
\begin{figure*}
\includegraphics[height=4in,width=6in]{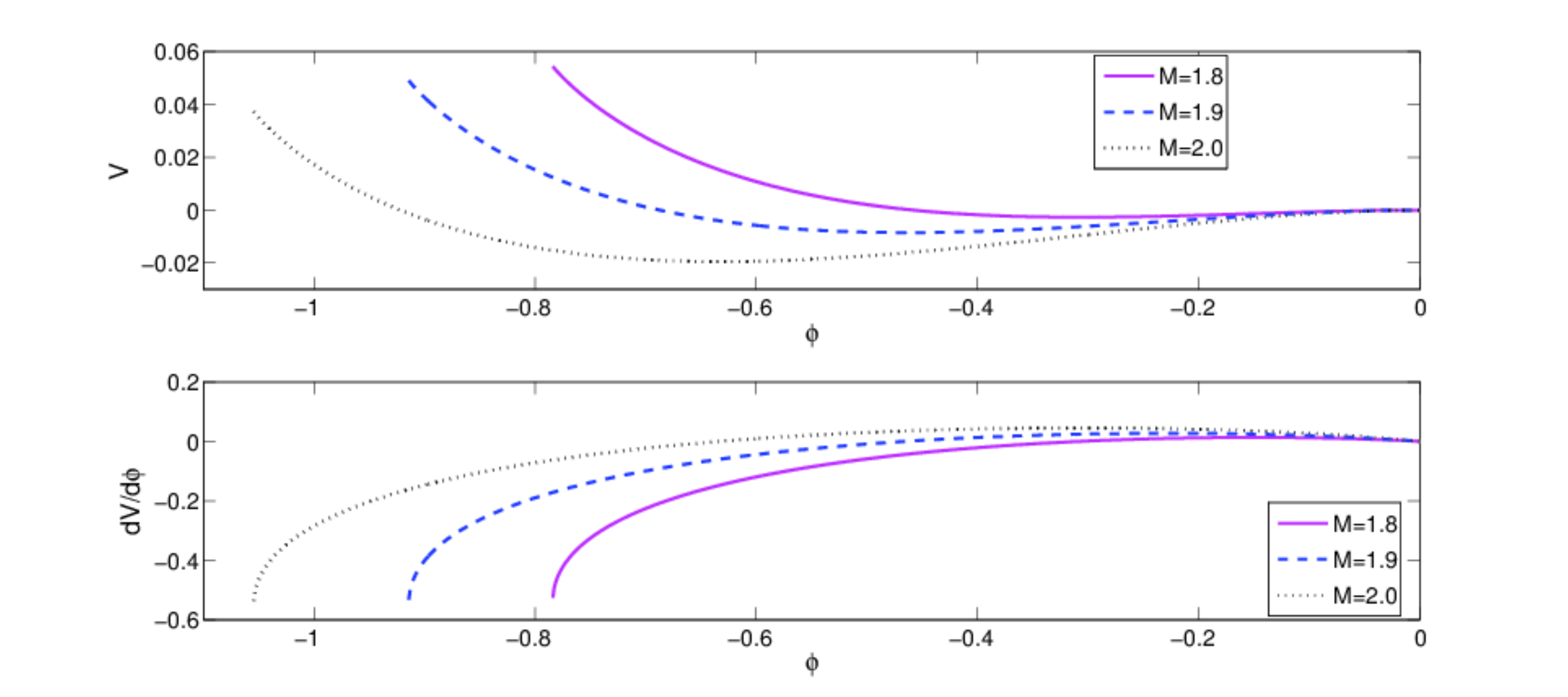}  
\caption{(Color online) The upper and lower panels, respectively, show the plots of the pseudopotential $V$ [Eq.\eqref{V-pseudo}] and its derivative $dV/d\phi$ [Eq. \eqref{dv-dphi}]   against the electrostatic potential $\phi$ for different values of the Mach number $M$. The solid, dashed and dotted   lines correspond to { $M=1.8$, $1.9$  and $2.0$ respectively. The other parameter values are $m=2.5$, $T=0.1$ and $\delta=0.1$.   In this figure, we see that $V=0=dV/d\phi$ and $d^2V/d\phi^2<0$ (not shown) at $\phi=0$. Also, $V=0$ and $dV/d\phi<0$ at $\phi=-0.458$ (solid line), $-0.68$ (dashed line) and $-0.92$  (dotted line). The Conditions 1 to 4 are satisfied for $1.8\lesssim M\lesssim2.1$ and for fixed parameters as above. Furthermore, $dV(\phi_m)/d\phi\neq0$ except at $\phi=0$, implying the nonexistence of   double-layer solutions.  }} 
\end{figure*} 
\begin{figure*}
\includegraphics[height=4in,width=6in]{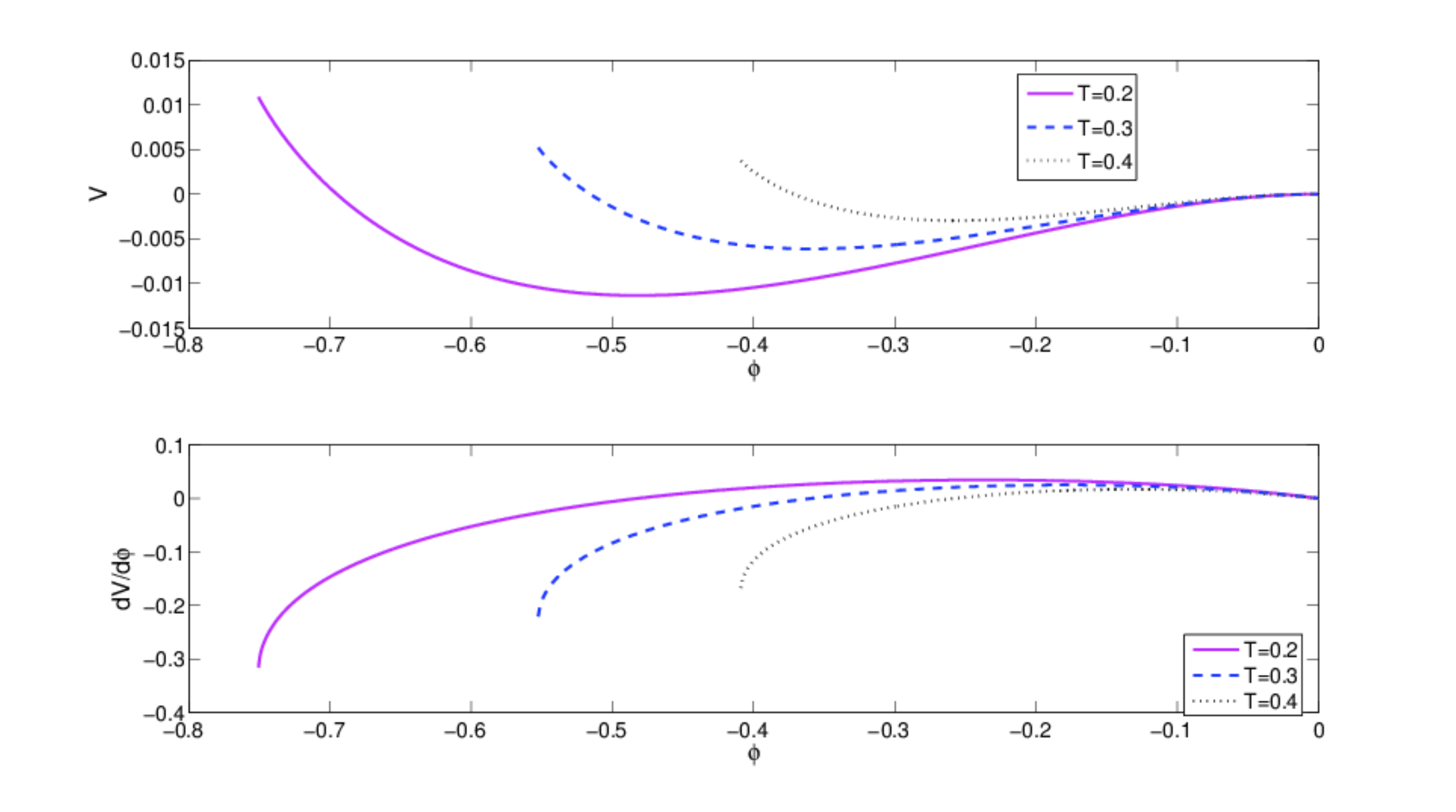}  
\caption{(Color online) The upper and lower panels, respectively, show the plots of the pseudopotential $V$ [Eq.\eqref{V-pseudo}] and its derivative $dV/d\phi$ [Eq. \eqref{dv-dphi}]   against the electrostatic potential $\phi$ for different values of the temperature ratio $T$. The solid, dashed and dotted   lines correspond to  { $T=0.2$, $0.3$ and $0.4$  respectively. The other parameter values are $m=2.5$, $M=2$ and $\delta=0.1$.  We find that $V=0=dV/d\phi$ and $d^2V/d\phi^2<0$ (not shown) at $\phi=0$. Also, $V=0$ and $dV/d\phi<0$ at $\phi=-0.695$ (solid line), $-0.515$ (dashed line) and $-0.37$  (dotted line). The Conditions 1 to 4 are satisfied for $T\lesssim0.7$ and for fixed parameters as above. Furthermore, $dV(\phi_m)/d\phi\neq0$ except at $\phi=0$, implying the nonexistence of   double-layer solutions. }} 
\end{figure*}
\begin{figure*}
\includegraphics[height=4in,width=6in]{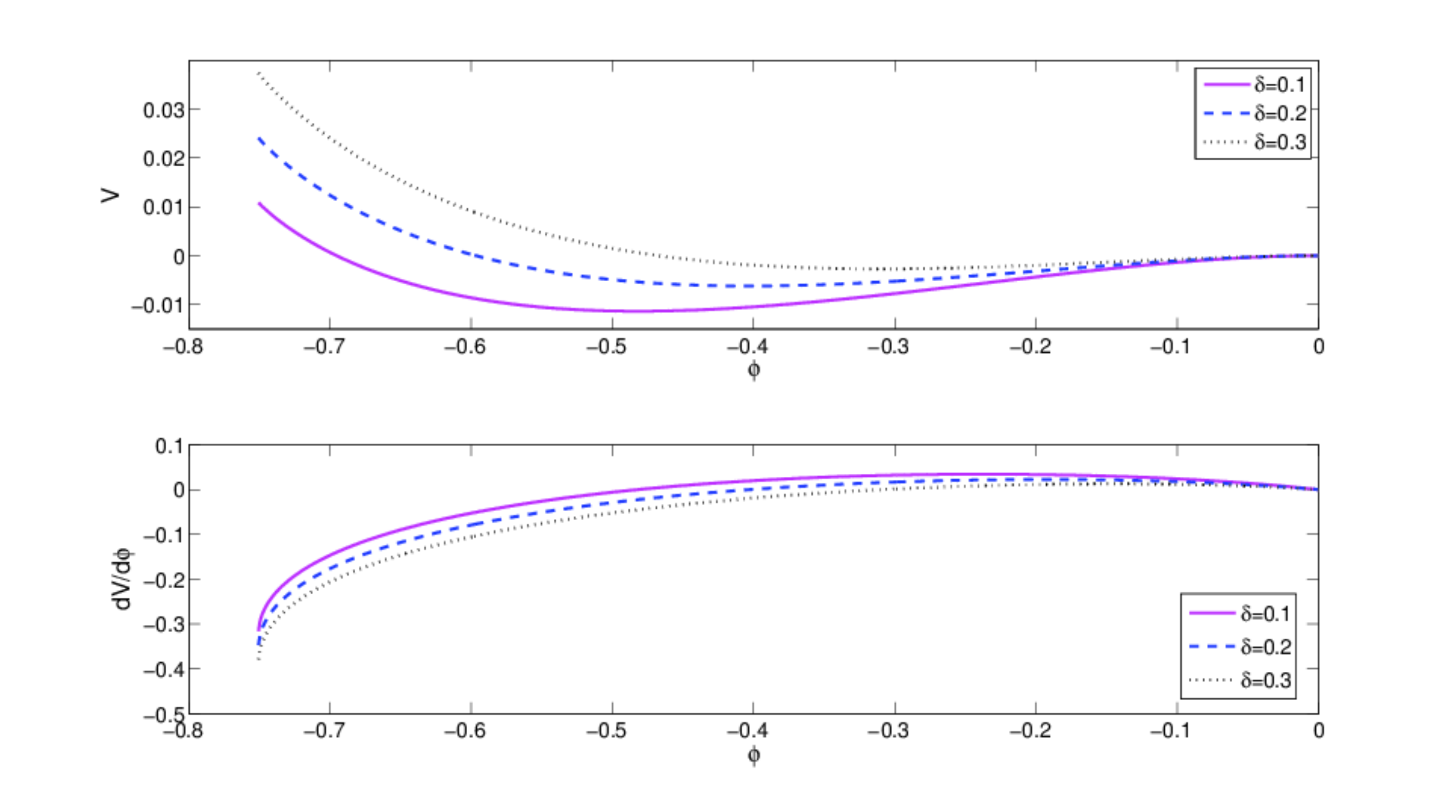}  
\caption{(Color online)  {The upper and lower panels, respectively, show the plots of the pseudopotential $V$ [Eq.\eqref{V-pseudo}] and its derivative $dV/d\phi$ [Eq. \eqref{dv-dphi}]   against the electrostatic potential $\phi$ for different values of the dust-to-negative ion density ratio $delta$. The solid, dashed and dotted   lines correspond to $\delta=0.1$, $0.2$ and $0.3$  respectively. The other parameter values are $m=2.5$, $M=2$ and $T=0.2$. Here, we see that $V=0=dV/d\phi$ and $d^2V/d\phi^2<0$ (not shown) at $\phi=0$. Also, $V=0$ and $dV/d\phi<0$ at $\phi=-0.695$ (solid line), $-0.603$ (dashed line) and $-0.47$  (dotted line). The Conditions 1 to 4 are satisfied for $\delta\lesssim0.35$ and for fixed parameters as above. Furthermore, $dV(\phi_m)/d\phi\neq0$ except at $\phi=0$, implying the nonexistence of   double-layer solutions.   }} 
\end{figure*}
Next, we  numerically investigate the  conditions 1 to 4 stated above for the existence SWs or double layers.  {Figures 4 to 7 show the profiles of the pseudopotential $V$ (upper panel) and its derivative $dV/d\phi$ (lower panel) for different values of the parameters $m$, $M$, $T$ and $\delta$. We find that for certain parameter values, $V$ crosses the $\phi$-axis and $dV/d\phi<0$ at only negative values of $\phi$ in $\phi_l<\phi<\phi_u$, implying the existence of only negative SWs.  These negative values   of $\phi$  represent the heights or amplitudes $\phi_m$ of the large-amplitude solitons.  The latter can   be obtained   by  numerically  solving Eq. \eqref{energy-equation}.  From Figs. 4 to 7, it is clear that there is no common value of $\phi$ $(\neq0)$ for which $V(\phi)=0$ and $dV/d\phi>0$  are satisfied for the existence of positive SWs.  Notice, however,  that there are also some  parameter regimes corresponding to $m$, $M$,  $T$ and $\delta$ for which either $V$ $(<0)$ does not cross the $\phi$-axis or $V(\phi)>0$, and hence no potential well is formed for particle trapping. For example, in Fig. 4,  $V(\phi\neq0)\neq0$   for   $m\lesssim1.8$. In this case, the large-amplitude solitary profiles may exist for $m\gtrsim1.8$ with a fixed set of parameters $M=2$ and $T=0.1=\delta$. The similar ranges   for $M$, $T$ and $\delta$ (See Figs. 5 to 7) for the existence of large-amplitude solitons, respectively, are  $1.8\lesssim M\lesssim2.1$ with   fixed   $m=2.5$ and $T=0.1=\delta$; $T\lesssim0.7$ with   fixed   $m=2.5$, $M=2$, and $\delta=0.1$; $\delta\lesssim0.35$ with  a fixed set  $m=2.5$, $M=2$, and $T=0.2$.   We find that the absolute value of the depth of the potential well decreases with increasing values of $m$ (Fig. 4), $T$ (Fig. 6)  and $\delta$ (Fig. 7), which may lead to the enhancement of the width  and detraction of the amplitude $|\phi_m|$ of the SWs. This width (amplitude) may, however, be decreased (increased) with increasing values of the Mach number $M$ as in Fig. 5.}  
 
Our next attempt is to  numerically solve Eq. \eqref{energy-equation} to exhibit the profiles of the large-amplitude  solitons. These  are shown in  { Fig. 8 for different sets of parameters $m$, $M$,  $T$ and $\delta$. Evidently, the amplitude of the large-amplitude soliton decreases while its width increases with increasing values of $m$ and $\delta$. On the other hand, as the values of $M$ $(T)$ increase, the soliton amplitude increases (decreases) while its width decreases.}  It may be interesting to examine the profiles of the SWs whose amplitudes tend to zero. Such small-amplitude solutions of the SWs can be obtained by expanding $V$ in powers of $\phi$ up to $\phi^3$ around the origin, i.e., $V(\phi)=C_1\phi^2+C_2\phi^3$, where $C_1=(1/2)\left(d^2V/d\phi^2\right)_{\phi=0}$ and $C_2=(1/6)\left(d^3V/d\phi^3\right)_{\phi=0}$, and using the boundary conditions, namely $\phi\rightarrow0$, $d\phi/d\zeta\rightarrow0$ as $\zeta\rightarrow\pm\infty$, as:
 \begin{equation}
 \phi=\phi_m\text{sech}^2\left(\zeta/W\right) \label{small-amp-sagdeev}.
 \end{equation}
Here, $\phi_m=-C_1/C_2$ is the amplitude and $W=\sqrt{4/|C_1|}$ is the width of the soliton. Alternatively, one can follow a perturbation  technique to investigate the dynamical evolution as well as the properties of KdV solitons. This will be done    in the next section.
 \begin{figure*}
\includegraphics[height=4in,width=6in]{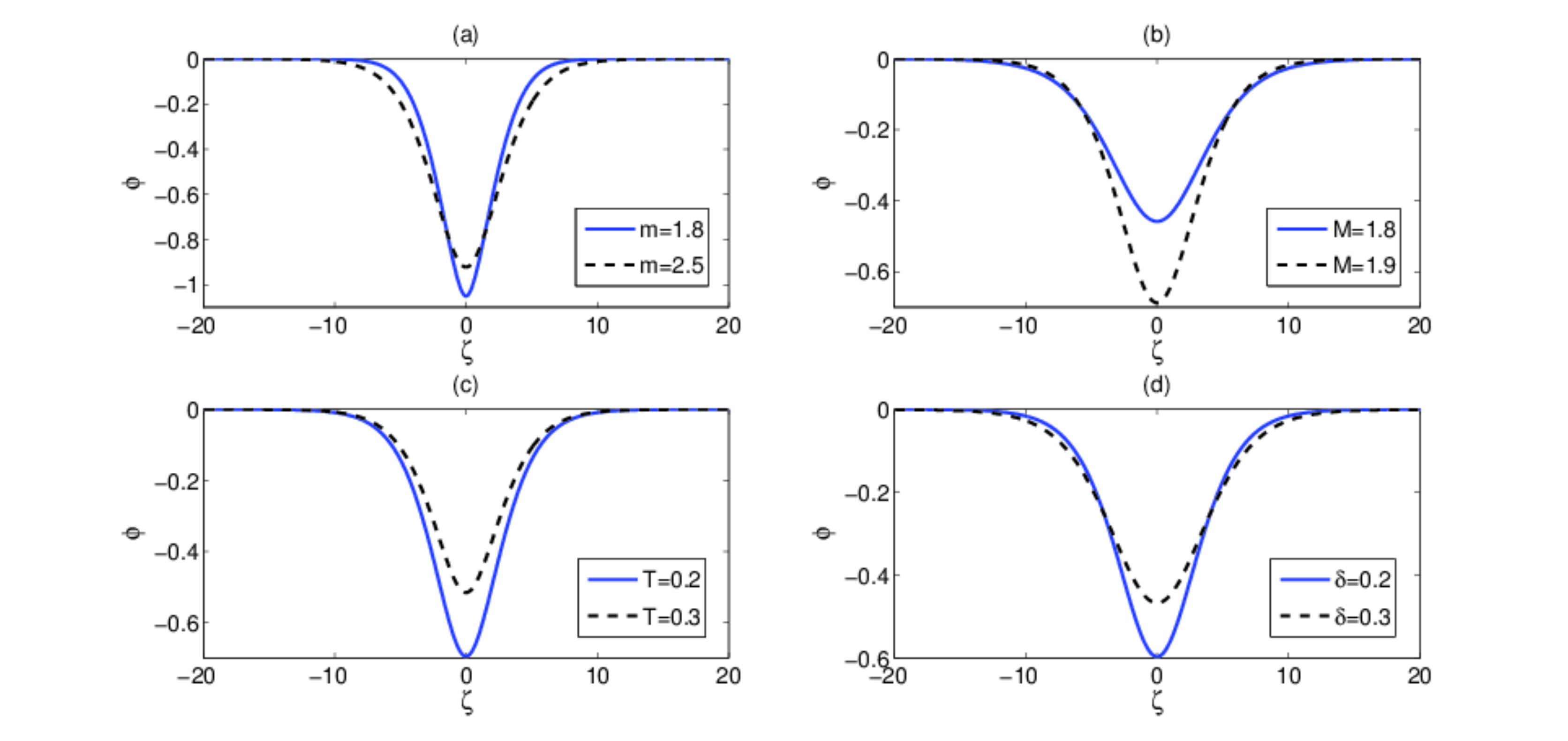}  
\caption{ (Color online)  {Large amplitude soliton solution of Eq. \eqref{energy-equation} is shown for different values of (a) $m$ (upper left), (b) $M$ (upper right), (c) $T$ (lower left) and (d) $\delta$ (lower right). The other parameter values are corresponding to  the Figs. 4 (dashed and dotted lines), 5 (solid and dashed lines), 6 (solid and dashed lines) and 7 (solid and dashed lines) respectively. It is seen that the  amplitudes of the solitons are exactly the same at which $V$ crosses the $\phi$-axis in between $\phi_l$ and $\phi_u$.}  } 
\end{figure*} 
\section{Small-amplitude KdV soliton: Perturbation technique}
We consider  the nonlinear propagation of   small but finite amplitude  electrostatic  waves in   dusty pair-ion plasmas.  In the standard reductive perturbation technique, the stretched coordinates are considered as $\xi=\epsilon^{1/2}(x-V_0t)$ and $\tau=\epsilon^{3/2}t$,  where $\epsilon$ is a small parameter $(\lesssim1)$ measuring the weakness of perturbations and $V_0$ is the wave speed.     The dynamical variables are expanded as  
\begin{eqnarray}
&&n_j=1+\epsilon n_j^{(1)}+\epsilon^2 n_j^{(2)}+\cdots,\notag \\
&&v_j=\epsilon v_j^{(1)}+\epsilon^2 v_j^{(2)}+\cdots,\label{expansion-nvphi} \\
&&\phi=\epsilon \phi^{(1)}+\epsilon^2 \phi^{(2)}+\cdots.\notag \\
\end{eqnarray}
We then substitute these expansions   and the stretched coordinates into   Eqs. \eqref{Ncont-eq}-\eqref{Npoisson-eq}, and equate different powers of $\epsilon$. In the lowest order of $\epsilon$ (i.e., $\epsilon^{3/2}$) we obtain the following relations for the first-order perturbations:
\begin{eqnarray}
&&\left( n_p^{(1)},v_p^{(1)} \right)=\left(1,V_0\right)\alpha\phi^{(1)}, \notag\\
&& \left(n_n^{(1)},v_n^{(1)}\right)=\left(1,V_0\right)\beta\phi^{(1)}, \label{1st-pertur} 
\end{eqnarray}
where $\alpha=m/(V_0^2-3m)$ and $\beta=1/\left(3T-V_0^2\right)$. The expression for the wave speed in the moving frame of reference is given by 
\begin{equation}
V_0=\sqrt{\frac{3m(1+\mu T)}{1+\mu m}}. \label{phase-velo}
\end{equation}
We note that since $m>T$ (for $m>1$ and $T<1$), we have  $\sqrt{3T}<V_0<\sqrt{3m}$ (in contrast to $M<\sqrt{3T}$ for large-amplitude SWs) and $V_0$ increases with increasing values of $m$ and $T$.
Proceeding to the next order of $\epsilon$ (i.e., $\epsilon^{5/2}$), the details are omitted for simplicity, we obtain a set of equations for the second-order perturbed quantities. The latter are then eliminated to obtain, after few steps, the following   KdV equation.
\begin{equation}
\frac{\partial\Phi}{\partial\tau}+A\Phi\frac{\partial\Phi}{\partial\xi}+B\frac{\partial^3\Phi}{\partial\xi^3}=0,\label{KdV-eq}
\end{equation}
where $\Phi\equiv\phi^{(1)}$. The coefficients of  nonlinearity and dispersion are,  respectively, given by
\begin{equation}
A=\frac{3\alpha\left[(1+m\mu^2)V_0^2+m(1+T\mu^2)\right]}{2V_0(1+m\mu)},\label{nonlinear-coeff}
 \end{equation}
\begin{equation}
B=\frac{m}{2V_0\mu\alpha^2(1+m\mu)}. \label{dispersion-coeff}
\end{equation}
Inspecting on the coefficients $A$ and $B$ [Eqs. \eqref{nonlinear-coeff} and \eqref{dispersion-coeff}], we find that  $B$ is always positive. Also,    $A$ is always negative, because $\alpha<0$ for  $V_0<\sqrt{3m}$. These imply that the small-amplitude SWs exist with only the negative potential.   The stationary soliton solution of the KdV equation \eqref{KdV-eq} can be obtained by applying a transformation $\eta=\xi-U_0\tau$, where $U_0$ is the constant phase speed normalized by $c_s$, and  imposing the boundary conditions for localized perturbations, namely, $\phi$, $d\phi/d\xi$, $d^2\phi/d\xi^2\rightarrow0$ as $\xi\rightarrow\pm\infty$ as:
 \begin{equation}
 \Phi=\Phi_m\text{sech}^2(\eta/w), \label{small-soliton}
 \end{equation}
 where $\Phi_m=3U_0/A$ is the amplitude and $w=\sqrt{4B/U_0}$ is the width of the soliton. Note that the qualitative behaviors, i.e., the properties of the amplitude and width of the  small-amplitude solitons given by Eqs. \eqref{small-amp-sagdeev} [Obtained from the energy-like equation  \eqref{energy-equation} with an approximation] and \eqref{small-soliton} [Obtained from the KdV equation \eqref{KdV-eq}]  for different values of $m$ and $T$ will remain the same. However, we analyze only the  properties of the KdV soliton.    Figures 9 and 10 show the profiles of the soliton amplitude (upper panel) and width (lower panel) [Eq. \eqref{small-soliton}] with the variation of  $\mu$ for different values of $m$ and $T$.  Figure 9 shows that the absolute value of the  amplitude decreases with $\mu$ and also with increasing values of $m$, whereas the width decreases with increasing $\mu$ (except for a fixed $m=1$ at which the width approaches a constant value) and  $m$. On the other hand,  the variations of the amplitude and width with $\mu$ for different values of $T$ show almost opposite features. In the latter, the amplitude (absolute value) increases but the width decreases with increasing values of $T$. In both the figures 9 and 10, the soliton amplitude is seen to decrease with $\mu$ and  approaches  more or less a constant value as $\mu$ approaches $1$.
  \begin{figure*}
\includegraphics[height=4in,width=6in]{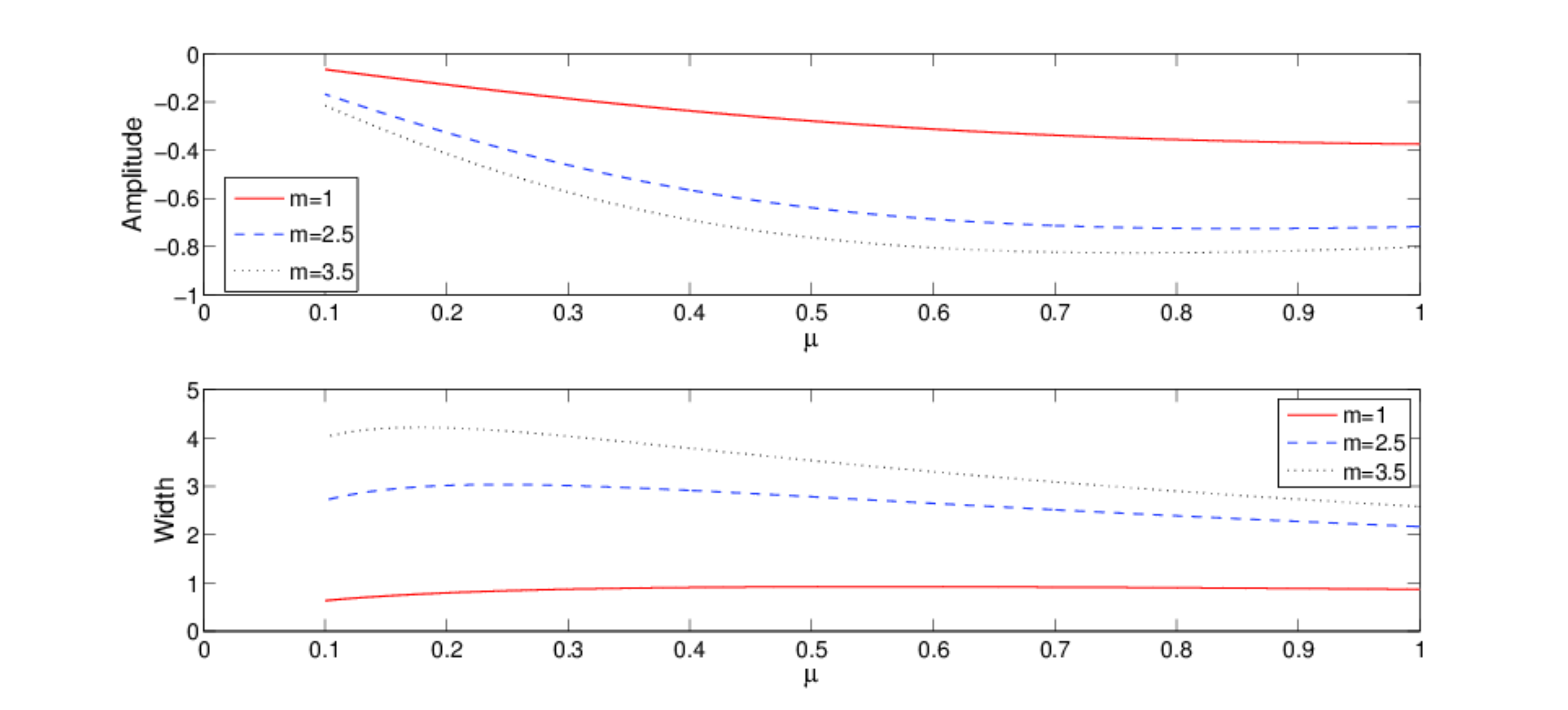}  
\caption{ (Color online)   The amplitude (upper panel) and width (lower panel) of the KdV soliton given by Eq. \eqref{small-soliton} are plotted against the density ratio $\mu$ for a fixed temperature ration $T=0.5$ and for different values of $m$: $m=1$ (solid line), $2.5$ (dashed line) and $3.5$ (dotted line).   } 
\end{figure*} 
  \begin{figure*}
\includegraphics[height=4in,width=6in]{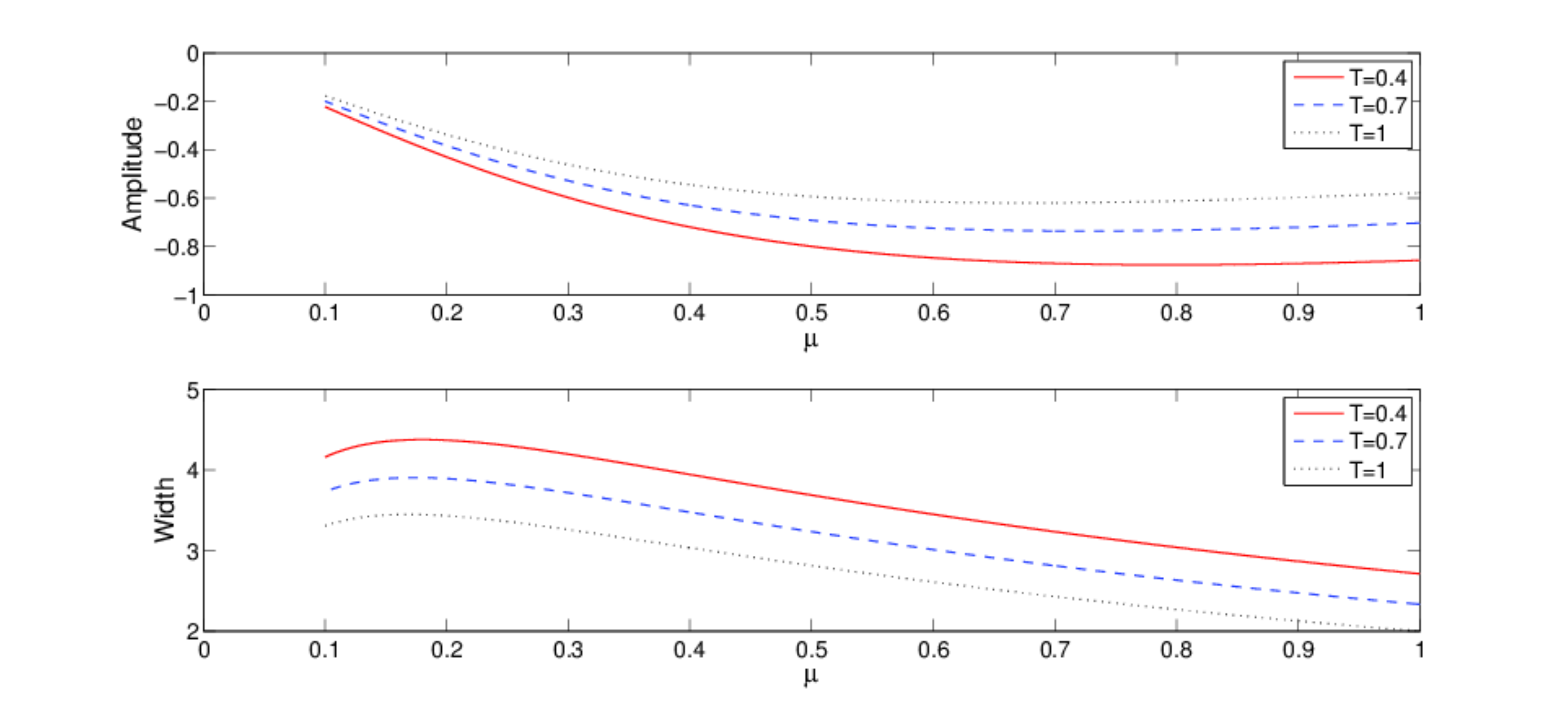}  
\caption{ (Color online)   The amplitude (upper panel) and width (lower panel) of the KdV soliton given by Eq. \eqref{small-soliton} are plotted against the density ratio $\mu$ for a fixed mass ration $m=3.5$ and for different values of $T$: $T=0.4$ (solid line), $0.7$ (dashed line) and $1$ (dotted line).  } 
\end{figure*} 
 
For the dynamical evolution of the soliton, we numerically solve Eq. \eqref{KdV-eq} by Runge-Kutta scheme with an initial condition  $\Phi=-0.05\exp\left[-(\xi+10)^2/100\right]$. The development of the pulse at different times  $\tau=0$ (upper left), $\tau=100$ (upper right), $\tau=200$ (lower left) and $\tau=400$ (lower right)  is shown in Fig. 11. The parameter values are considered as $m=4.5$, $T=0.6$ and $\delta=0.1$. It is seen that the leading part of the initial pulse steepens due to positive nonlinearity, and  as   time goes on, the pulse separates into solitons and a residue due to the wave dispersion.  Furthermore,  once the solitons are formed and separated, they propagate without changing their shape due to the nice balance of the nonlinearity and dispersion (see the plot for $\tau=400$).
  \begin{figure*}
\includegraphics[height=4in,width=6in]{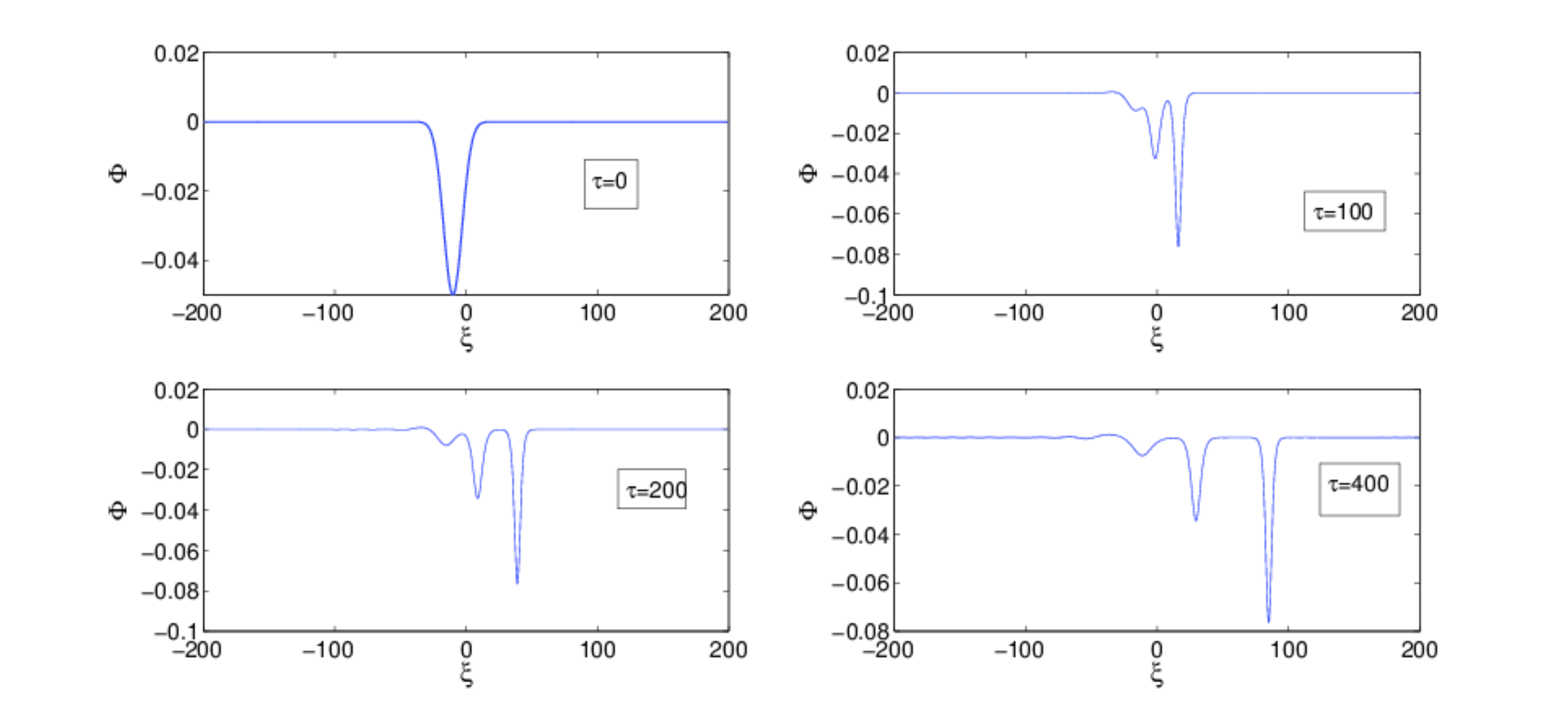}  
\caption{ (Color online)   The development of an initial pulse of the form $\Phi(\xi)=-0.05\exp\left[(\xi+10)^2/100\right]$ into solitary waves at different times $\tau=0$, $\tau=100$, $\tau=200$ and $\tau=400$ as in the figure. The parameters are considered as $m=4.5$, $T=0.6$ and $\delta=0.1$.    } 
\end{figure*} 

\section{Discussion and Conclusion}
 We have investigated the propagation characteristics of electrostatic waves in an unmagnetized collisionless pair-ion plasma with immobile positively charged dusts. The latter become positively charged when the number density of negative ions exceeds that of positive ions and is much larger than that of electrons  so that electron contribution can be neglected \cite{Kim-Merlino}. In the linear regime,  there appear two ion modes, namely ``fast" and ``slow" waves in dusty pair-ion plasmas.  Due to the higher values of the phase velocity of the fast ion wave than the ion thermal speeds, the Landau damping effect   is negligibly small. We find that in the long-wavelength limit $(k\rightarrow0)$, the fast (slow) mode propagates with a frequency greater (lower) than the frequency of   negative-ion oscillations.  Furthermore, in the limit $k\rightarrow0$, the frequency of the fast modes almost assumes a constant value.  We also find that the thermal pressure  has very small effect on the fast-wave modes, whereas it can modify the frequency of the slow-wave modes. In contrast to the effects of different mass of the ions,  the effect of positively charged dusts is to decrease the wave frequency of the fast waves.    

In the nonlinear regime, we have  studied the  propagation of large- as well as small-amplitude  perturbations in  dusty pair-ion plasmas. We show that perturbations can develop into solitary waves, and no double-layer solution can be formed. Using the pseudopotential approach, we derive a energy-like equation, which (along with the pseudopotential) is numerically analyzed to study the properties of the large-amplitude solitons for different values of the system parameters $m$, $T$ and $\mu$ or $\delta$ which satisfy  $m>1$, $T<1$ and $\mu<1$.  It is found that large-amplitude SWs exist for the  {Mach number satisfying $\text{max}\left\lbrace  \sqrt{3T},~\tilde{M}\right\rbrace<M<\sqrt{3m}$. In our numerical investigation,  the parameter regimes for the existence of large-amplitude SWs are obtained as (i) $m\gtrsim1.8$ with fixed $M=2$ and $T=0.1=\delta$, (ii)   $1.8\lesssim M\lesssim2.1$ with  a fixed set   $m=2.5$ and $T=0.1=\delta$, (iii) $T\lesssim0.7$ with   fixed   $m=2.5$, $M=2$, and $\delta=0.1$, and (iv) $\delta\lesssim0.35$ with  a fixed set  $m=2.5$, $M=2$, and $T=0.2$. It is shown that the effects of  different masses  $(m)$, different temperatures $(T)$ of ions as well as the charged dust impurity $(\delta)$ are to diminish the soliton amplitudes. The latter can be  enhanced by increasing the nonlinear wave speed $M$. Furthermore, the   widths of the large-amplitude solitons can be  increased (decreased) by the effects of $m$ and $\delta$ ($M$ and $T$). }  For small-amplitude waves, we derive a    KdVB equation which is numerically solved to present the dynamical evolution of solitons. Furthermore, we have numerically studied the properties of the amplitude and width of the KdV solitons for different values of the system parameters. We find that both the large- and small-amplitude waves  propagate with only the negative potential  in pair-ion plasmas with positively charged dusts.  The theoretical results may be useful for the observation of electrostatic waves in space plasmas, e.g., a dusty meteor trail region in the upper atmosphere, in industrial electron-free pair-ion plasmas as well as for the experimental verification of the excitation of ion-acoustic   waves in laboratory dusty pair-ion plasmas.  

\section*{acknowledgments} This work was partially supported by  the SAP-DRS (Phase-II), UGC, New Delhi, through sanction letter No. F.510/4/DRS/2009 (SAP-I) dated 13 Oct., 2009, and by the Visva-Bharati University, Santiniketan-731 235, through Memo No. Aca-R-6.12/921/2011-2012 dated 14 Feb., 2012.



\end{document}